\documentclass{caosp308}

\usepackage{graphicx}

\usepackage{natbib}
\articleNo{305}
\pubyear{2021}
\volume{51}
\volnumber{2}
\firstpage{138}
\received{November 5, 2020}
\accepted{February 25, 2021}

\begin{document}

\htitle{MASTER OT J172758.09+380021.5}
\hauthor{E.\,Pavlenko, T.\,Kato, K.\,Antonyuk, N.\,Pit, L.\,Keir, S.\,Udovichenko, 
P.\,Dubovsk\'{y}, A.\,Sosnovskij, O.\,Antonyuk, 
V.\,Shimansky, M.\,Gabdeev, F.\,Rakhmatullaeva, G.\,Kokhirova, S.\, Belan, 
A.\,Simon and A.\,Baklanov, N.\,Kojiguchi, V.\,Godunova
	 }
\hauthor{E.\,Pavlenko {\it et al.}}

\title{MASTER OT J172758.09+380021.5: a peculiar ER UMa-type dwarf nova, probably a missed nova in the recent past}

\author{
        E.\,Pavlenko\inst{1}  
      \and
        T.\,Kato\inst{2}
      \and   
        K.\,Antonyuk\inst{1}   
      \and 
        N.\,Pit\inst{1}
      \and
       L.\,Keir\inst{3}
       \and
       S.\,Udovichenko\inst{3}
       \and
       P.\,Dubovsk\'{y}\inst{4}
       \and
       A.\,Sosnovskij\inst{1}
       \and
       O.\,Antonyuk\inst{1}
       \and
       V.\,Shimansky\inst{5,6}
       \and
       M.\,Gabdeev\inst{5,6,7}
       \and
       F.\,Rakhmatullaeva\inst{8}
       \and
       G.\,Kokhirova\inst{8}
       \and
       S.\,Belan\inst{1}
       \and 
       A.\,Simon\inst{9}
       \and
       A.\,Baklanov\inst{1}
       \and
       N.\,Kojiguchi\inst{2}
       \and
       V.\,Godunova\inst{10}
             }

\institute{
           \ Federal State Budget Scientific Institution ``Crimean Astrophysical Observatory of RAS'', 
           Nauchny, 298409, Republic of Crimea, \email{eppavlenko@gmail.com}
         \and
           \ Department of Astronomy, Kyoto University, Kyoto 606-8502, Japan   
         \and 
           \ Astronomical Observatory, I. I. Mechnikov Odessa National University, Odessa oblast, Ukraine 
         \and 
           \ Vihorlat Observatory, Mierova 4, 06601 Humenne, Slovakia
         \and
           \ Kazan (Volga region) Federal University, Kazan, 420008,\\ Kremlyovskaya 18, Russia
         \and
            \ Special Astrophysical Observatory, 369167, Nizhnij Arkhyz, Karachai-Cherkessian Republic, Russia  
         \and
            \ Tatarstan Academy of Science Institute of Advanced Studies, 4221, Levobulachnaya St. 36 A, Kazan, Tatarstan republic, Russia
         \and
           \ Institute of Astrophysics of the Academy of Sciences of the Republic of Tajikistan, Bukhoro 22, Dushanbe, 734042, Tajikistan
         \and
           \ Astronomy and Space Physics Department, Taras Shevshenko National University of Kyiv, 
           Volodymyrska str. 60, Kyiv, 01601, Ukraine  
         \and
           \ ICAMER Oservatory of NAS of Ukraine, 27 Acad. Zabolotnogo str., 03143~Kyiv, Ukraine
          }
      
\date{November 5, 2020}

\maketitle

\begin{abstract}
A CCD photometry of the dwarf nova MASTER OT J172758.09 +380021.5 was carried out in 2019 during 134 nights. Observations covered three superoutbursts, five normal outbursts and quiescence between them. The available  ASASSN and ZTF  data for 2014 -- 2020 were also examined. Spectral observations were done in 2020 when the object was in quiescence. Spectra and photometry revealed that the star is an H-rich active ER UMa-type dwarf nova with a highly variable supercycle (time interval between two successive superoutbursts) of $\sim$ 50 -- 100 d that implies a high and variable mass-transfer rate.  MASTER OT J172758.09+380021.5 demonstrated peculiar behaviour: short-lasted superoutbursts (a week); a slow superoutburst decline and cases of rebrightenings; low frequency (from none to a few) of the normal outbursts during the supercycle. In 2019 a mean period of positive superhumps was found to be 0.05829 d during the superoutbursts. Late superhumps with a mean period of 0.057915 d  which lasted about $\sim$ 20 d after the end of superoutburst and were replaced by an orbital period of 0.057026 d  or its orbital-negative superhump beat period were detected. An absence of eclipse in the orbital light curve and its moderate amplitude are consistent with the orbital inclination of about 40$\degr$ found from spectroscopy. The blue peaks of the \textit{V-Ic} and \textit{B-Rc} colour indices of superhumps during the superoutburst coincided with minima of the light curves, while \textit{B-Rc} of the late superhumps coincided with a rising branch of the light curves. We found that a low mass ratio $q = 0.08$ could explain most of the peculiarities of MASTER OT J172758.09+380021.5. The mass-transfer rate should be
accordingly higher than what is expected from gravitational
radiation only, this assumes the object is  in a post-nova
state and underwent a nova eruption relatively recently  -- hundreds of years ago. This object would provide probably the first observational evidence that a nova eruption can occur even in CVs near the period minimum.
\end{abstract}
\vspace{-5.mm}

\section{Introduction}
\label{intr}
Cataclysmic variables (CVs) are close binary systems in which a white dwarf (WD) accretes matter from a late-type donor star that fills its Roche lobe. Accretion is either disc-like or pole-on if a WD is non-magnetic or magnetic one, respectively. The orbital periods of CVs are limited by a rather blurry border of about 12 hours and by a strong period minimum at $\sim$78 min \citep{2001cvs..book.....H}. There is a so-called “orbital period gap” between 2.15 and 3.18 hours \citep{2006MNRAS.373..484K}  with a significant deficit of non-magnetic CVs. Non-magnetic CVs in the period gap and below it up to the period minimum are the SU UMa-type stars \citep{1995CAS....28.....W}. These binaries are the H-rich stars approaching during their evolution to the period minimum or passed it ("period bouncers"),  further evolution of period bouncers takes place with an increase of the orbital period. 56 system are known at present which have shorter periods than the period minimum. They are the He-rich  AM CVn stars \citep{2018MNRAS.476.1663G} evolving with a decrease of the period. CVs on a way to AM CVn stars are called "EI Psc-type stars" \citep{2016PASJ...68...65K}, they exhibit both H and He and are located near the period minimum. Another subgroup of SU UMa-stars, namely WZ-type stars,  could also be found close to the period minimum \citep{2015PASJ...67..108K}. 
Transfer of matter from the late-type companion leads to thermal instability of the accretion disk and causes its outburst - a dwarf nova (DN) event \citep{1996PASP..108...39O}. SU UMa stars display two types of outbursts: the normal outbursts that last 2-5 days and brighter and longer superoutbursts lasting up to several weeks.

During the superoutbursts periodic brightness variations (positive superhumps) with periods of a few percent longer than the orbital period are present. These superhumps, as it was shown by   \cite{1988MNRAS.232...35W}; \citet*{1989PASJ...41.1005O}; \citet{1990PASJ...42..135H}; \citet{1991ApJ...381..268L}, are the consequence of tidal instability resulting from the 3:1 resonance in the accretion disc.
This can only happen for the SU UMa-type stars with a mass ratio $q=M_{2}/M_{1} \le 0.3$, where $M_{2}$ and $M_{1}$ are the masses of the donor and WD stars, respectively.
 
  \citet{2009PASJ...61S.395K}  introduced three stages in the positive superhump evolution: Stage A has longer superhump periods and growth of superhump amplitudes; superhump periods during Stage B could systematically decrease/increase or be stable with a decrease of their amplitudes; Stage C has shorter superhump periods. 

Some of the DNe display so-called “late superhumps” \citep{1983A&A...118...95V}. They have a similar period \textbf{as} positive superhumps, but have phases shifted by a half of this period. One explanation of this phenomenon is that they arise from  a hot spot on an elliptical disc \citep{1985A&A...144..369O}.

Contrary to the positive superhumps, the negative superhumps have periods a few percents shorter than the orbital one. Negative superhumps mostly appear in quiescence independently on the mass ratio \citep{2010AIPC.1273..358M, 2019cwdb.confE..39P} and probably are caused by nodal precession of a tilted accretion disk \citep{2003cvs..book.....W}. \citet*{2014PASJ...66...15O} suggested that decrease of the frequency of the normal outbursts in some DNe resulting in a transition from a short (S) cycle to a long (L) one (designations of cycles, i.e., the interval between two successive outbursts, is taken from \citet{1985AcA....35..357S}) could be caused by a transition of a disc from coplanar with an orbital plane to a tilted to the orbital plane and the subsequent appearance of nodal precession and negative superhumps. This suggestion was confirmed for NY Her \citep{2017IBVS.6216....1S}, V1504 Cyg \citep{2013PASJ...65...95O}  and V503 Cyg \citep{2019cwdb.confE..39P}.

WZ-type stars have in average a lowest mass transfer rate and rare outbursts that occur every several years –- decades and actually are the superoutbursts with no normal outbursts with rare exceptions \citep{2015PASJ...67..108K}.   One of the outstanding features of WZ Sge-type stars is rebrightenings, which are outbursts that appear after the plateau of the superoutburst during the superoutburst decline \citep{2015PASJ...67..108K}. WZ-type stars display a long-lasting approach to the quiescence. The superhumps during this stage (late-stage superhumps) could be observable, indicating that the disc in this stage is still thermally and tidally unstable.

About 25 years ago a special sub-group of SU UMa stars, ER UMa-type stars \citep{1995PASJ...47..163K, 1995PASJ...47..897N},   was distinguished among the DNe as having an extremely highest frequency of superoutbursts (every 20-50 days). 

DN MASTER OT J172758.09+380021.5 (hereafter  MASTER 1727) was discovered by \citet{2014ATel.5724....1D}  in an outburst at $V=14^{m}.3$.  \citet{2016AJ....152..226T} obtained spectrum in quiescence at $V\sim 18^{m}.5$. It showed very strong, relatively narrow single-peaked emission lines of H. The emission-line velocities indicated an orbital period of 82.14(6) minutes. The authors predicted the superhumps detection in the nearest future.  Indeed, superhumps were discovered in the 2019 superoutburst   (see vsnet-alerts 22236 and 23335)\footnote{https://groups.yahoo.com/neo/groups/cvnet-outburst/}, where the superhump period was reported to be 0.0565(1) d and 0.05803(1) d, respectively. K. Naoto previously suggested that this object might be the EI Psc-type candidate  (vsnet-alert 23720). According to the available ASASSN \citep{2014ApJ...788...48S} light curve\footnote{http://cv.asassn.astronomy.ohio-state.edu/} one could suggest a rather short supercycle (interval between two successive superoutbursts) of $\sim 50-60$ d  and unusually low frequency of normal outbursts between them within JD 2458415 -- 2458650.

In this paper, we present a spectral and photometric study of MASTER 1727  at different stages of its activity based on our observations in 2019 -- 2020 and on the ASASSN and ZTF\footnote{https://lasair.roe.ac.uk/} \citep{2019PASP..131a8003M}  database in 2014 -- 2019. 

\section{Observations and data reduction}

Spectral observations were {carried} out at the 6-m telescope of SAO RAS on 29 April 2020 using {the} SCORPIO reducer with a long slit mode. {We used the
volume holographic grating with 1200 lines per mm (VPHG1200G) and the 1" slit. We obtained the	spectra in the range $\lambda\lambda = 3900-5700$\,{\angstrom}  with the resolution of 5.5 \angstrom. Five consecutive exposures with a total duration of 25 minutes were obtained. For the wavelength and flux calibration we used the $HeNeAr$ lamp and a spectrophotometric standard. The observed	data reduction was performed with standard methods under the IDL\footnote{http://www.ittvis.com/idl} environment.} Bright sky background, light clouds and seeing of stars about 2.5" provided the resulting S/N ratio $\sim$ 40. 

Photometric observations were carried out in 2019 during 134 nights (174 independed runs of observations) from June 25 to November 5 in the Crimean Astrophysical Observatory, Vihorlat Observatory, and in Observatories of Mayaki, Sanglokh and Terskol (see the log of observations given that is available in an electronic form). Most of the observations were done in white light (without any filter) but several observations close to the Johnson-Cousins \textit{B, V, Rc} and \textit{Ic} system were obtained. We used the comparison star 1280-0316078 of the USNO B1.0 catalogue. Depending on astroclimatic conditions (i.e. sky background, light absorption in the atmosphere and seeing), telescope and brightness of the object, exposure time  varied and provided $S/N = 20 - 200$. We used the MAXIM DL package for aperture photometry after standard data processing (i.e., debiasing, dark subtracting, flat-fielding).

For further analyses of the time series we used the Stellingwerf method implemented in the ISDA package \citep{1980faat.book.....P} and used  the O-C analysis for the times of light curves maxima. For this and analysis of multicolour observations, we used the MCV program \citep{2004AstSR...5..264A}.


\section{Results}
\subsection{Spectrum in the 2020 quiescence}

\begin{figure}
	\centerline{\includegraphics[width=0.75\textwidth,clip=]{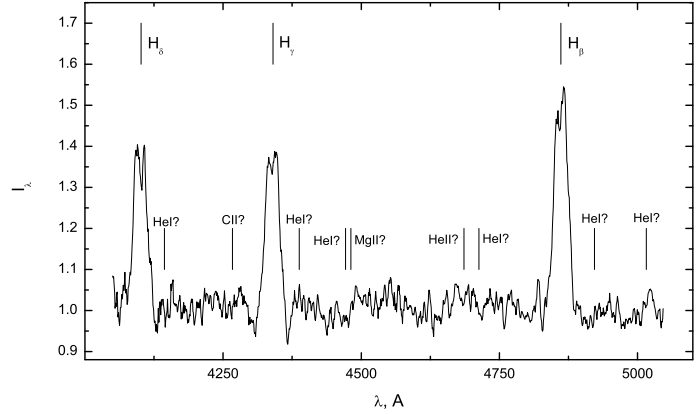}}
	\caption{Spectrum of MASTER 1727 in the 2020 quiescence.}
	\label{sp1}
\end{figure}

\begin{figure}
	\centerline{\includegraphics[width=0.7\textwidth,clip=]{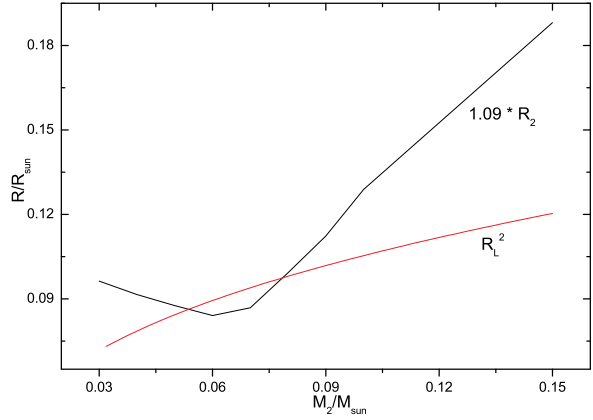}}
	\caption{Dependences of the secondary star radius $ R_ {2} $ and its Roche-lobe radius $ R_ {L} ^ {2} $ on its mass $ M_ {2}$.}
	\label{figur1}
\end{figure}

In the spectrum, double-peaked emission lines of the Balmer series with a $FWHM = 23 - 27$ $\angstrom$ are observed (see Fig. \ref{sp1}). The lines of neutral and ionized He are not found with exception of a weak emission near 5015 \angstrom, which probably is an artefact of sky lines subtraction. Besides of these lines, the CII and Mg II ions are also absent in the spectrum. Peak-to-peak distances in the HI ($\Delta \lambda = 13.2-13.9$ \angstrom) correspond to rotational velocity of the outer radius of the accretion disc $Vsin(i)=410-450$ km/s. Assuming the WD Roche-lobe radius $R_{L} = 0.4R_{\sun}$, its mass  $M_{1} = 0.75M_{\sun}$ and the ratio of the outer radius of the accretion disk to the Roche-lobe size equal to 0.8, one could estimate the orbital inclination $i = 40\degr \pm 2\degr$.  

According to the theoretical calculations of   \citet{1977ApJ...216..822P}, the size of this radius is limited by the tidal effect of the secondary component and is about 0.8 $R_{L}^{1}$. To find the value of $R_{L}^{1}$ we applied the method for estimating the masses of the components of the close binaries developed by  \citet{2017AstBu..72..184B}.   The WD mass corresponds to $M_{1}=0.75M_{\sun}$,  the average mass of the primary component WD
 \citep{2011yCat....102018R}. For a large set of mass ratios of the components  $q = M_{2}/M_{1}$, we calculated the mass of the secondary component $M_{2}$ and the size of its Roche lobe according to the Eggleton formula \citep{1983ApJ...268..368E}:
\begin{equation}
\label{r1} 
R_{L}^{2}={{0.49q^{2/3}} \over {0.6q^{2/3}+\ln(1+q^{1/3})}}, \ \ \ \ 0<q<\infty
\end{equation} 

We estimated the radius of the secondary component $R_{2}$ for each value of its mass based on the results of evolutionary calculations of  \citet{2003A&A...402..701B} and 
 \citet{2000A&AS..141..371G} for the low-mass dwarfs. The resulting dependences of $ R_ {2} $ and $ R_ {L} ^ {2} $ on $ M_ {2} $ were compared   (see Fig. \ref{figur1}) taking into account the assumption that for systems with stable accretion the following condition is valid: 
\begin{equation}
\label{r2}
R_{L}^{2} = 1.05-1.09 \sim R_{2}
\end{equation}

For $M_{2}=0.070M_{\sun}\pm 0.005M_{\sun}$,  ensuring the fulfillment of the condition, we calculated by the formula \citep{1983ApJ...268..368E} Roche-lobe radius of the primary component  $R_{L}^{1} = 0.347R_{\sun}\pm{0.030}R_{\sun}$  and the outer radius of the accretion disk  $R_{AD} = 0.28R_{\sun}\pm0.03R_{\sun}$. Comparing the Keplerian velocity at this radius   $V = 690$ km/s  with the observed values presented above  we can find the angle of inclination of the orbit is  $i = 41{\degr}\pm{5}\degr$. 

\subsection{A long-term light curve}

In this subsection, we consider supercycles, cycles and morphology of superoutbursts based on our data obtained in 2019 and on the available ASASSN and ZTF data obtained in 2014 -2019.

\subsubsection{2019 light curve}

\begin{figure}
	\centerline{\includegraphics[width=0.75\textwidth,clip=]{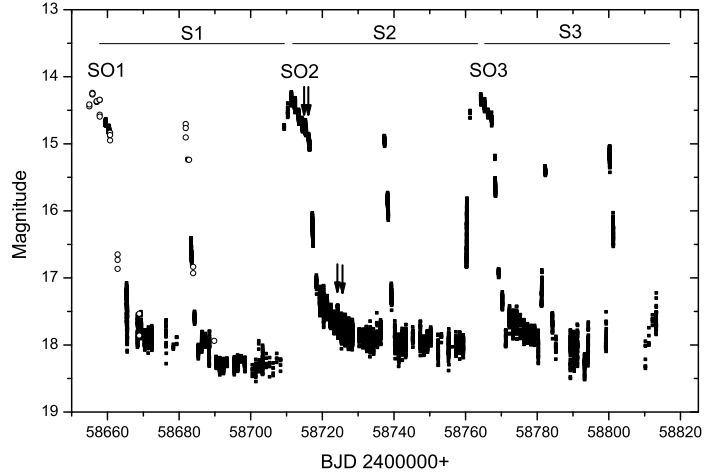}}
	\caption{An overall 2019 light curve of MASTER 1727. SO1, SO2 and SO3 denote the number of superoutbursts while S1, S2 and S3 denote the number of supercycles. 
		Our and ASASSN data are marked by black and open circles,  respectively. Arrows indicate the time of colourimetric observations.}
	\label{Fig1}
\end{figure}
The overall 2019 light curve of Master 1727 is shown in Fig. \ref{Fig1}. It is constructed using our and part of ASASSN  data. It covered three superoutbursts with the amplitude of $\sim 4^{m}$  and four normal outbursts. All superoutbursts displayed an unusually short duration of about one week. This is shorter than a range of typical   H-rich DNe \citep{1995CAS....28.....W} with exception of the 6-d duration in RZ LMi \citep{1995PASP..107..443R}, though there are a few  He-rich AM CVn systems \citep{2019AJ....157..130C} with a duration of superoutbursts less than 10 days. The superoutburst plateau of MASTER 1727+38 declined with a rate of $\sim$ 0.136 mag/d. Then the superoutburst displayed rapid decay of $\sim$2.5 mag in two days,  subsequent slow return to quiescence that lasted up to the beginning of the next superoutburst, and a short supercycle of 52 d.
 The normal outbursts lasted about two days. There was only one normal outburst just in the middle between superoutbursts, this activity repeated for two neighbour supercycles. However, the next supercycle included at least two normal outbursts. The appearance of outbursts during a  slow approaching of superoutbursts to quiescence looks like rebrightenings often seen in the WZ Sge-type stars \citep{2015PASJ...67..108K}.

\subsubsection{ASASSN and ZTF 2014 -- 2019 light curve}

\begin{figure}
		\centerline{\includegraphics[width=0.75\textwidth,clip=]{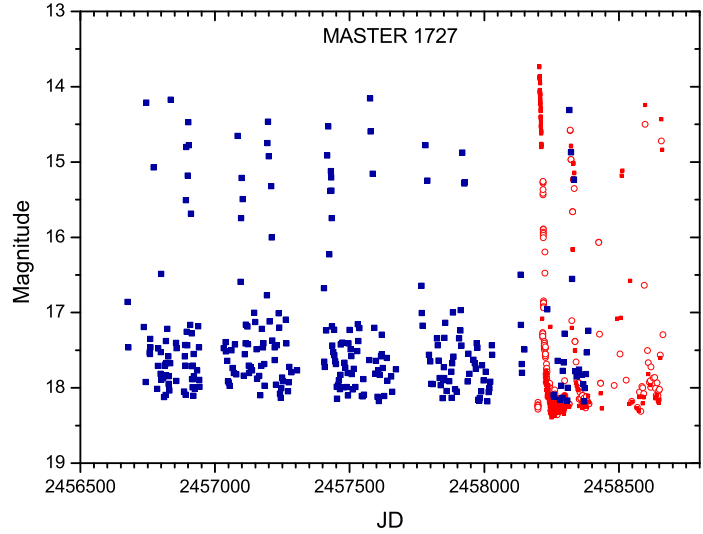}}
	\caption{The $\sim$5.5-year light curve. ASASSN and ZTF data are shown by blue and red colours, respectively.}
	\label{LC-long1}
\end{figure}

\begin{figure}
	\centerline{\includegraphics[width=0.75\textwidth,clip=]{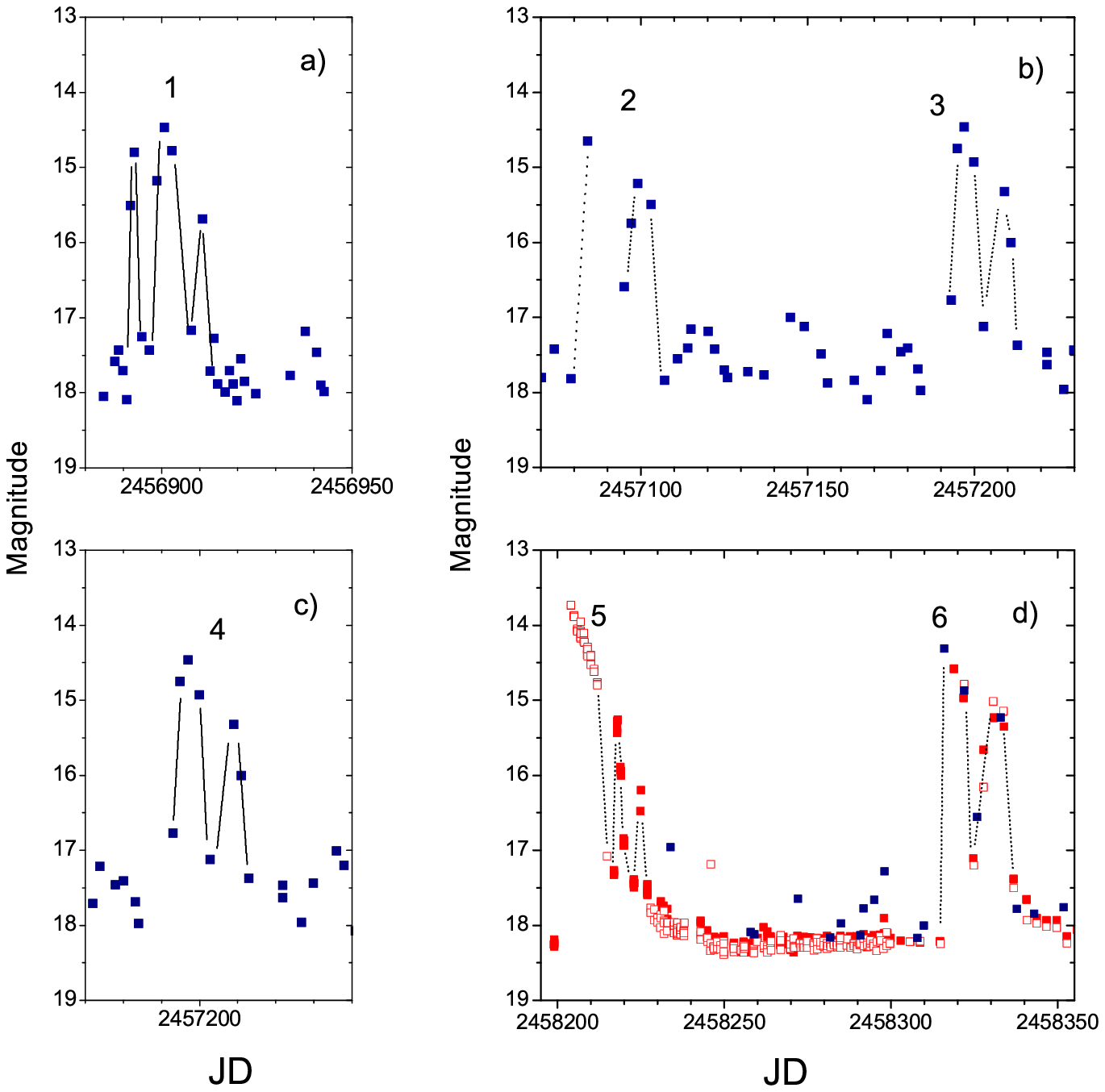}}
	\caption{Fragments (a, b, c, d) of the long-term light curve displaying complex morphology of the outburst profile. The first event probably associates with a normal outburst + superoutburst + rebrightening, the second -- fourth events and the sixth one with the superoutburst + rebrightening and the fifth event with the superoutburst and two rebrightenings. ASASSN and ZTF data are shown by blue and red colours, respectively.}
	\label{events}
\end{figure}
It was possible to explore the outburst activity of MASTER 1727 on the $\sim$5.5-year scale (2014 -- 2019) using the available ASASSN and ZTF data (see Fig. \ref{LC-long1}). Contrary to our observations, we cannot immediately identify a type of all bright events (whether they are normal outbursts or superoutbursts) for this data. Moreover, in the case of the the superoutburst, maximum of brightness potentially could refer to any moment within the 7-day plateau. Probably all the data of $14^{m} - 14.5^{m}$ refer to the superoutbursts.
 The impression from the visual inspection of the data is that the supercycle is not constant on a 5.5-year scale. Due to the short duration of the superoutburst, risk of missing it in a not very dense observation series makes it difficult to accurately determine the supercycle length. Nevertheless, denser observations (also taking into account our data) allow us to conclude that the supercycle could change in a range of 52 -– 110 d during 1.5 years at least in 2018 -– 2019. Note that no normal outbursts occurred during the 110-d supercycle.
Of all the data on a 5.5-year scale, six events with a complex profile can be distinguished (Fig.\ref{events}). The most striking among them are the  fifth and sixth events. The fifth event obviously is the superoutburst with two rebrightenings. A plateau of this superoutburst lasted a $\sim$week. The sixth event could be understood as a superoutburst lasting 6 -- 9 d and the subsequent rebrightening with an unusually long ($\sim$6 d) rising branch.  The first -- fourth events probably can also be related to superhumps with rebrightenings.

\subsection{Superhumps during 2019 SO2 superoutburst}
	
\begin{figure}
	\centerline{\includegraphics[width=0.95\textwidth,clip=]{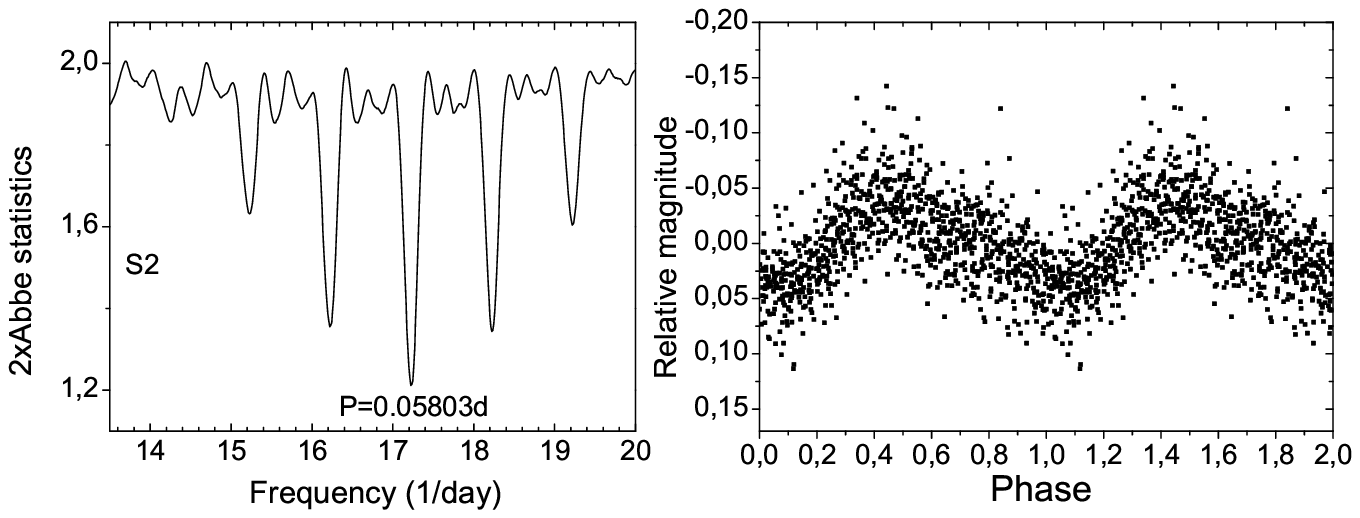}}
	\caption{Left: a periodogram for the S2 superoutburst plateau in the vicinity of the superhump period. A corresponding zero epoch is 2458711.29888. The most significant peak indicates the period of 0.05803(4) d. Right: data folded on this period. }
	\label{sh}
\end{figure}

Due to the fact that the SO2 superoutburst plateau was closely covered by observations, we calculated its mean superhump period. A periodogram indicates the most significant period of 0.05803(4) d among the one-day aliased peaks (see Fig. \ref{sh})  that we attribute to the positive superhump period. Its mean the profile is asymmetric one with an amplitude of about $0^{m}.1$. Over a course of all superoutbursts and in more detail during the SO2 superoutburst  (Fig.\ref{fig4a}),  the original amplitudes of superhumps decrease from $0^{m}.15$ to $0^{m}.05$  This means that superhumps were already fully grown since the superoutburst maxima (stage B) and stage A was probably too brief  (a day or so) to be catched by us.

\begin{table}[t]
	\small
	\begin{center}
		\caption{Times of nightly light curves maxima of the MASTER J1727 in 2019.}
		\label{t2}
		\begin{tabular}{lllllllr}
			\hline\hline
			
			BJD & Error & BJD & Error & BJD & Error & BJD & Error \\
			2458000+&  &  2458000+  & &  2458000+ & &  2458000+ & \\
			\hline
			
			659.457 & 0.003 & 715.212 & 0.003 & 722.318 & 0.004 & 733.324 & 0.003\\ 
			659.517 & 0.003 & 715.270 & 0.002 & 722.324 & 0.003 & 736.332 & 0.005\\  
			660.333 & 0.004 & 715.330 & 0.005 & 722.325 & 0.002 & 736.343 & 0.004\\ 
			660.395 & 0.003 & 715.331 & 0.002 & 722.381 & 0.002 & 738.264 & 0.005\\  
			660.456 & 0.002 & 715.332 & 0.003 & 722.441 & 0.003 & 743.298 & 0.006\\  
			665.356 & 0.004 & 715.389 & 0.003 & 723.252 & 0.002 & 745.245 & 0.007\\ 
			665.419 & 0.003 & 715.445 & 0.004 & 723.311 & 0.003 & 747.293 & 0.005\\  
			668.377 & 0.003 & 716.144 & 0.003 & 723.375 & 0.003 & 748.328 & 0.006\\  
			668.432 & 0.003 & 716.202 & 0.003 & 724.298 & 0.002 & 749.230 & 0.004\\  
			669.415 & 0.003 & 716.316 & 0.002 & 724.359 & 0.002 & 749.290 & 0.005\\  
			670.400 & 0.003 & 716.317 & 0.003 & 724.414 & 0.002 & 749.325 &0.006\\	
			670.456 & 0.003 & 716.318 & 0.005 & 725.279 & 0.002 & 752.311 & 0.003\\ 
			671.380 & 0.008 & 716.375 & 0.003 & 725.279 & 0.003 & 755.317 & 0.004\\ 
			671.503 & 0.005 & 716.376 & 0.002 & 725.336 & 0.002 & 759.315 & 0.003\\ 
			672.400 & 0.003 & 716.435 & 0.003 & 725.390 & 0.004 & 760.297 & 0.005\\ 
			683.319 & 0.01  & 717.160 & 0.003 & 725.396 & 0.002 & 764.296 & 0.004\\ 
			683.445 & 0.008 & 717.225 & 0.004 & 726.319 & 0.003 & 765.213 & 0.003\\ 
			686.347 & 0.005 & 717.396 & 0.003 & 727.255 & 0.002 & 766.265 & 0.005\\ 
			686.410 & 0.007 & 718.317 & 0.003 & 727.310 & 0.008 & 770.243 & 0.005\\ 
			691.390 & 0.01  & 719.300 & 0.004 & 727.314 & 0.004 & 770.295 & 0.005\\ 
			710.346 & 0.002 & 719.301 & 0.003 & 727.368 & 0.008 & 772.288 & 0.004\\ 
			711.331 & 0.002 & 719.421 & 0.003 & 727.373 & 0.004 & 772.261 & 0.004\\ 
			712.311 & 0.002 & 720.292 & 0.004 & 728.293 & 0.004 & 773.263 & 0.004\\ 
			712.311 & 0.003 & 720.295 & 0.003 & 728.296 & 0.004 & 774.251 & 0.002\\ 
			712.370 & 0.005 & 720.356 & 0.005 & 728.347 & 0.004 & 776.276 & 0.005\\ 
			713.360 & 0.005 & 720.360 & 0.003 & 728.352 & 0.003 & 778.241 & 0.002\\ 
			713.412 & 0.003 & 720.361 & 0.003 & 730.253 & 0.003 & 779.232 & 0.002\\
			714.288 & 0.003 & 721.280 & 0.003 & 730.269 & 0.004 & 781.246 & 0.003\\
			714.342 & 0.005 & 721.281 & 0.002 & 731.309 & 0.003 & 782.185 & 0.003\\
			714.343 & 0.004 & 721.332 & 0.003 & 731.368 & 0.003 & 782.223 & 0.004\\
			714.398 & 0.002 & 721.334 & 0.003 & 731.368 & 0.003 & 791.258 & 0.005\\
			714.402 & 0.003 & 721.394 & 0.005 & 732.344 & 0.002 & 800.220 & 0.005\\
			714.460 & 0.003 & 722.268 & 0.002 & 732.400 & 0.003 & 801.192 & 0.005\\
				\hline\hline
	\end{tabular}
	\end{center}
\end{table}

The periodic light variations with a period close to a superhump period continued over the quiescence between superoutbursts during every supercycle. 
The times of maxima for nightly light curves were defined wherever possible (see Tab.\,\ref{t2}) and (O-C)s were calculated using the  
zero epoch of BJD 2458710.346 and the period of 0.0580294 d.
The O-C for the superhumps maxima of the SO2 superoutburst displayed a possibly slight increase of the mean period.

\subsection{Periodicity in quiescence}

\begin{figure}
	\centerline{\includegraphics[width=0.75\textwidth,clip=]{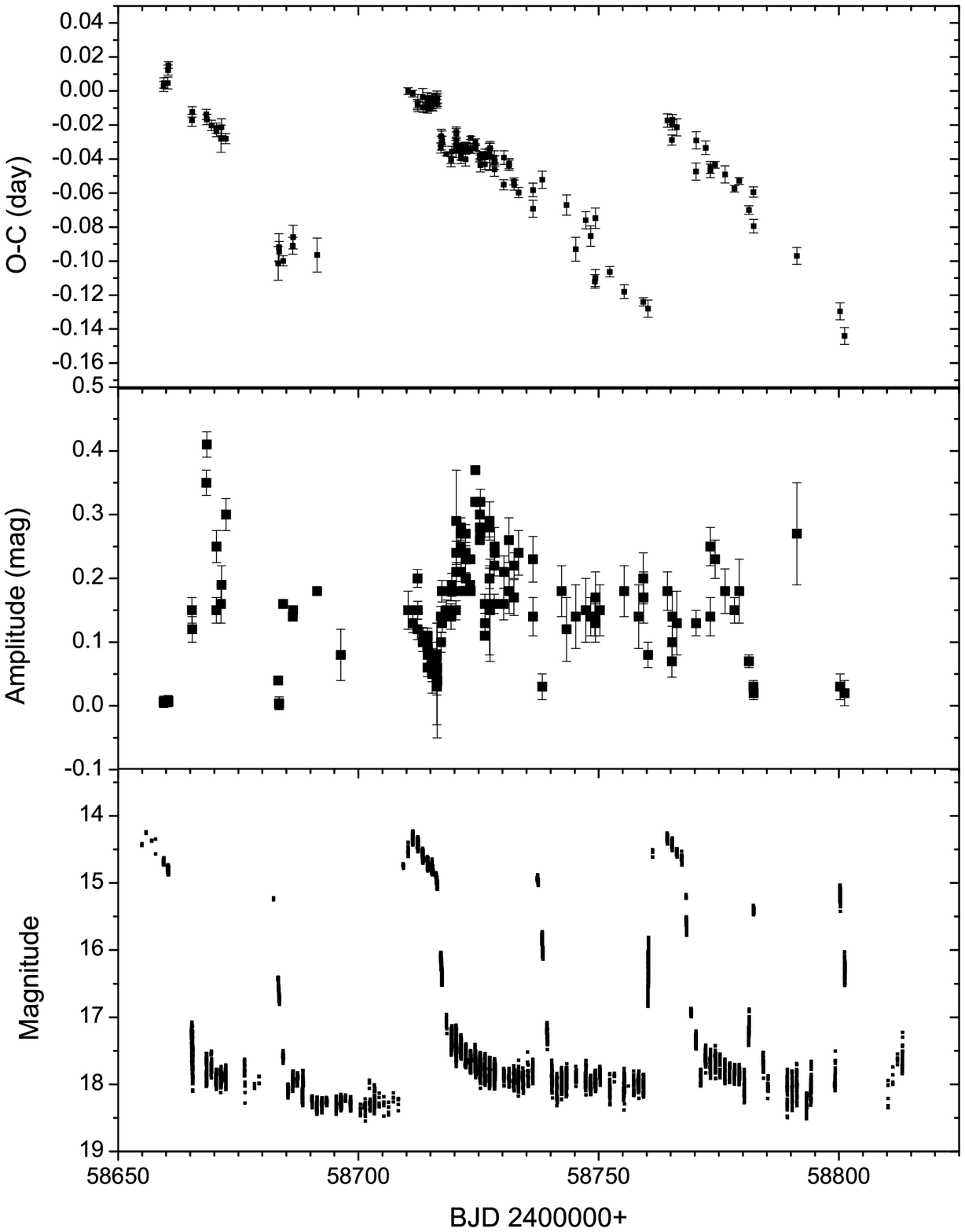}}
	\caption{ From top to bottom: O-C for maxima; amplitudes and an overall light curve during three S1, S2 and S3 supercycles.}
	\label{fig3a}
\end{figure}

\begin{figure}
	\centerline{\includegraphics[width=0.70\textwidth,clip=]{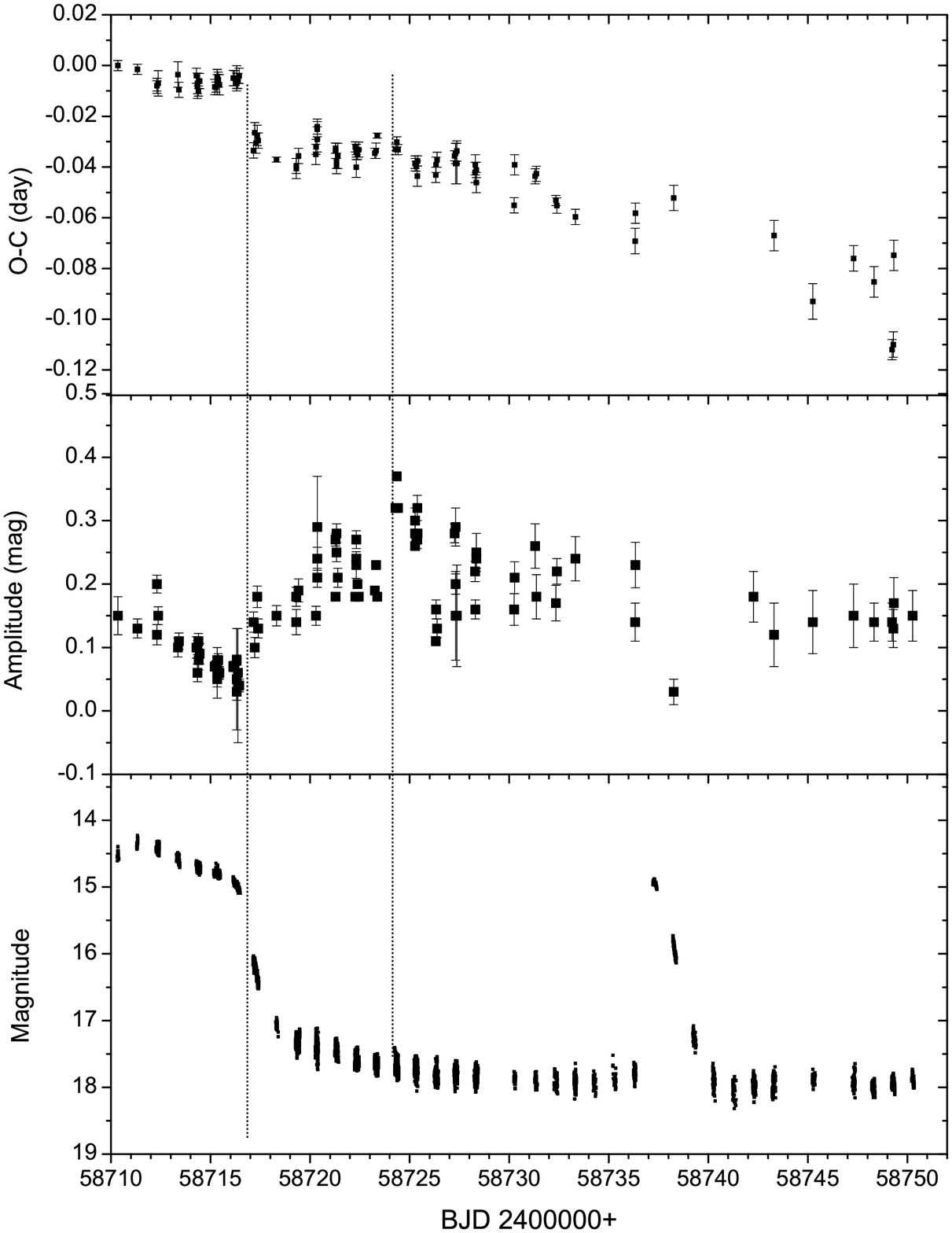}}
	\caption{The same as in Fig. \ref{fig3a}, but for the part of the S2 supercycle only.}
	\label{fig4a}
\end{figure}

Starting with BJD 2458717, which was $\sim$ a middle of the fast SO2 superoutburst decline, the (O-C)s displayed a clear jump by a half the period of positive superhumps. This event looked like classical late superhumps \citep{1983A&A...118...95V, 2002A&A...395..541K, 2002PASJ...54..599U}  that is shown in Fig.\ref{fig3a} and Fig.\ref{fig4a}. Further (O-C)s  decreased up to the next superoutburst that means that the corresponding period was shorter than superhumps in the superoutburst.  Simultaneously the mean amplitude of this periodicity grew from $0^{m}.05$ to $0^{m}.35$ up to the onset of the normal outburst.  After that, the cycle-to-cycle amplitudes varied around the mean amplitude of  $\sim$  $0^{m}.15$ up to the end of the supercycle. 
  

\begin{figure}
 	\centerline{\includegraphics[width=0.78\textwidth,clip=]{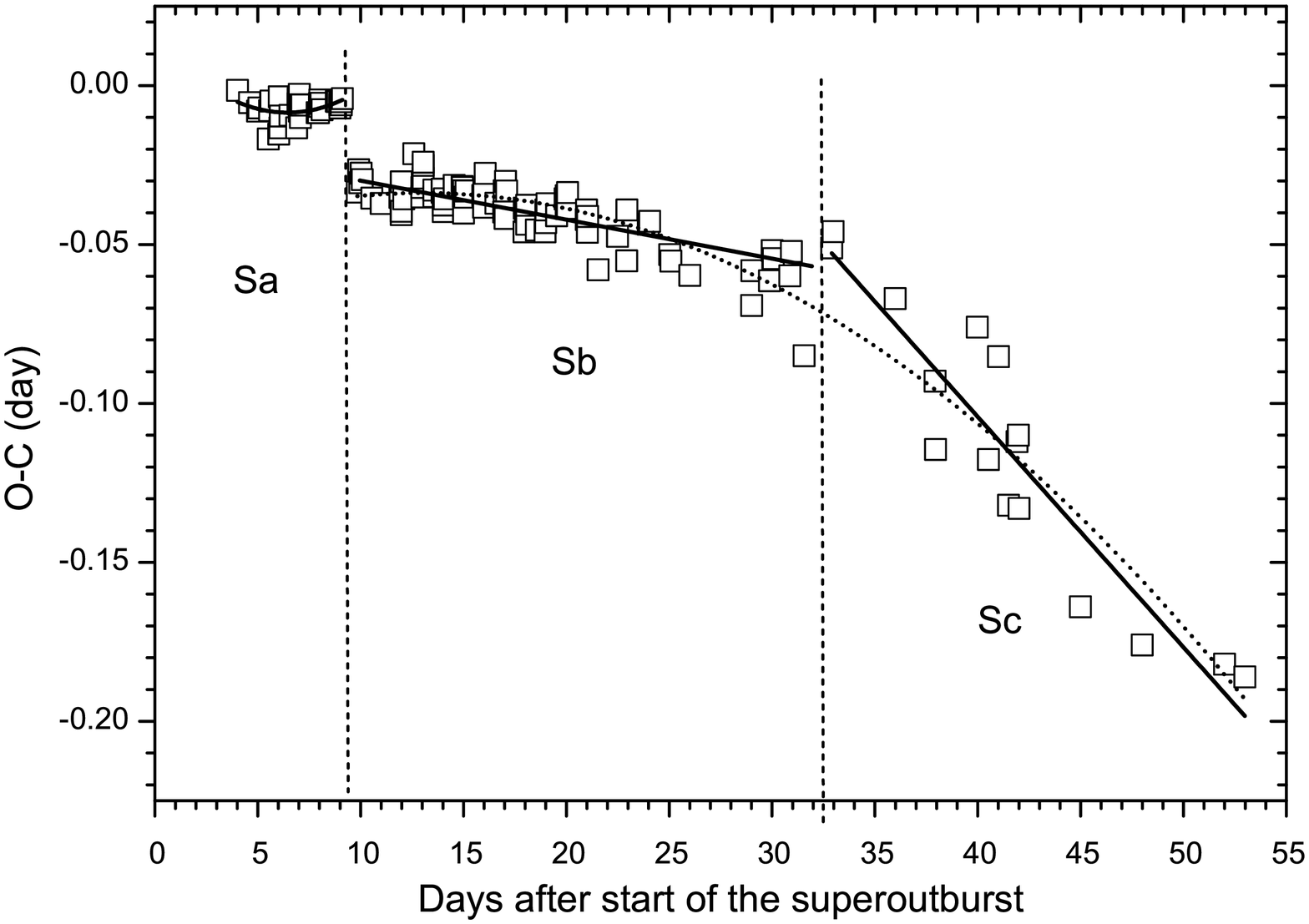}}
 	\caption{O-C combined for three S1, S2 and S3 supercycles.  Sa, Sb and Sc mean three stages of combined supercycles: Sa means a plateau of the superoutburst, Sb and Sc are the parts of supposed late superhumps. The dashed curve is the best parabolic fit of (O-C)s behaviour at $Sb + Sc$ stages; solid lines are the best fits of (O-C)s at Sb and Sc stages separately.}
 	\label{Fig5-a}
 \end{figure}

 It seems that the O-C behaviour repeated from supercycle to supercycle. To study the nature of periods in quiescence we combined the O-C for all S1, S2 and S3 supercycles by choosing the zero-point as the onset of the corresponding superoutburst. Despite poorer statistics for S1 and S3 supercycles, the O-C behaviours are consistent with each other. We have identified three stages in the O-C behaviour: a stage of the superoutburst plateau Sa, quiescence Sb that is limited by the end of the superoutburst and the normal outburst,  and quiescence Sc between the end of the normal outburst and the next superoutburst (see Fig. \ref{Fig5-a}). Limited and uneven coverage of the superhump plateau does not enable us to define the evolution of superhumps during stage B.  The mean period at this stage is 0.058029 d.  The stages Sb and Sc could be fitted by the linear trends which refer to two periods of 0.057938 d and 0.057611 d, respectively. Alternatively, Sb+Sc stages could be approximated by a parabola.  The scattering around linear fits seems to be slightly less than around the parabola, so the version of two periods in quiescence is more reliable.
 
\begin{figure}
	\centerline{\includegraphics[width=0.85\textwidth,clip=]{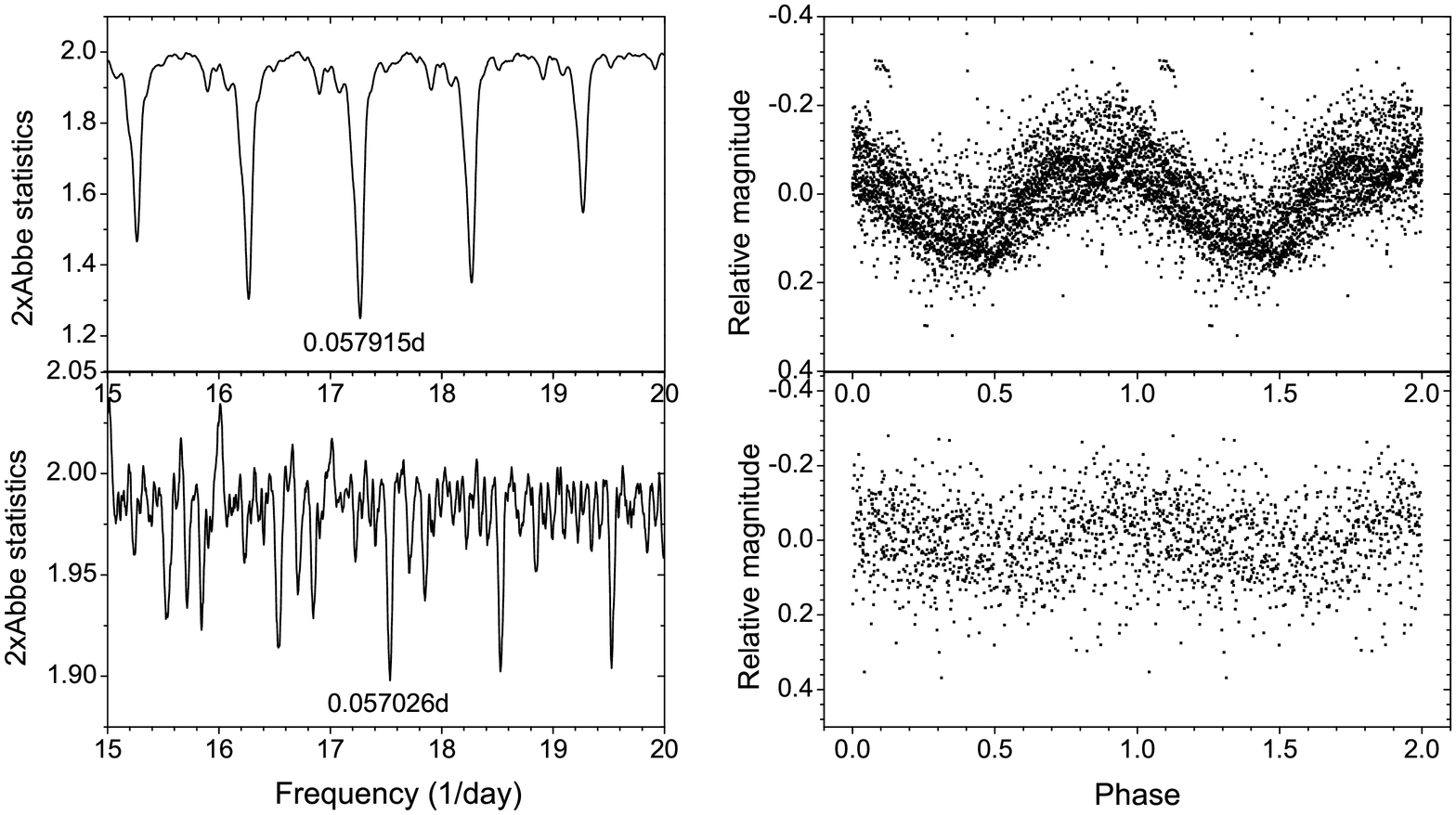}}
	\caption{Left from top to bottom: periodograms for data in the S2 quiescence before the normal outburst and after it;  Right from top to bottom: data folded with the most significant periods 0.057915 d and zero epoch 2458725.253;  0.057026 d and zero epoch 2458740.252, respectively.}
	\label{Fig6}
\end{figure}

We performed a periodogram analysis for the data at Sb and Sc stages using data of the S2 supercycle only (Fig. \ref{Fig6}).  The periods at these stages were 0.057915(10) d and 0.057026(10) d, respectively. One could see a good agreement of the O-C and periodogram analyses for the period at the Sb stage. The mean period at the stage Sb is 0.00011 d less than the mean period at the Sa stage. We suggest that this is a period of late superhumps that lasted about 20 d up to the start of a normal outburst. Note that  IY UMa \citep{2000PASP..112.1567P}  also displayed the period of late superhumps that was slightly shorter than during the superoutburst plateau. 
As for the Sc stage, the O-C also yields confirmation of the period obtained from the periodogram analysis but with less accuracy (probably because of an insufficient number of the O-C data). We adopt the 0.057026(9) d for the orbital period. 
 
According to \citet{2001cvs..book.....H} an expected fractional  period deficit $\epsilon^{-}$  of negative superhumps is $\sim$ $0.5\times \epsilon^{+}$ (here $\epsilon^{+}$ is a fractional period excess). In our case  $\epsilon^{-}$ should be $\sim$ -0.0087 and the period of negative superhumps has to be about 0.0565 d (frequency 17.70 $d^{-1}$). No prominent peak at this frequency  for both the superoutburst and the quiescence was found. There is only a week signal at this frequency during the stage C that we cannot take into account due to its low significance.  Instead, a period of 0.057026(9) d within errors coincides with the orbital period 0.057042(42) d  that was obtained from spectroscopy \citep{2016AJ....152..226T}. The mean light curve does not contain the eclipse and has a relatively low amplitude (about $0^{m}.1$). This does not contradict the moderate inclination of the orbit ($i = 40\degr \pm 2\degr$) that we obtained from spectroscopy.
 
\begin{figure}
	\centerline{\includegraphics[width=0.88\textwidth,clip=]{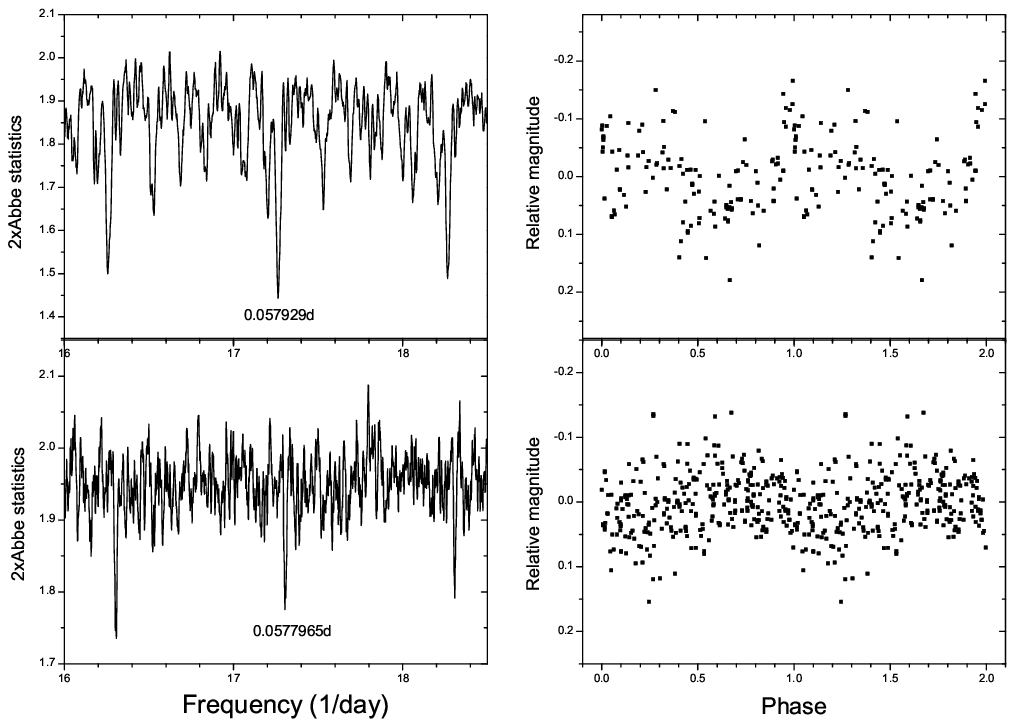}}
	\caption{Left from top to bottom: a periodogram for data of a slow superoutburst 6 decline and the subsequent quiescence; right from top to bottom: data folded on the periods of 0.057929 d and 0.0577965 d, respectively.}
	\label{ztf2a}
\end{figure}

Despite the insufficiently dense rows of the ASASSN and ZTF data for each night, a large number of nights in the quiescent state following the fifth superoutburst made it possible to search for periodicities in this quiescence. We have done this for data of a slow decline (JD 2458227 -- 2458250) and quiescence (JD 2458253 -- 2458315) separately.   A result is presented in Fig.\ref{ztf2a}.

A periodogram for the slow superoutburst decline revealed a most significant period of 0.057929(7) d. It coincides, within accuracy, with the period of late superhumps 0.057915(10) d found in 2019. Among the day-aliased peaks of periodogram for quiescent data one peak points to a period of 0.057797(3) d that is close to but is slightly less then the period of the late superhumps. In the case of our observations in 2019, where the second half of the supercycle was presented by orbital variations, the ZTF periodogram does not show this period. Possibly a period of late superhumps acted in quiescence and was somewhat smaller than during a slow fading, or there was cross-contamination of the late superhumps, orbital periodicity, potential negative superhumps or some of them (the negative superhumps could be  good candidates to an additional periodicity in the quiescence taking into account that this supercycle does not contain normal outbursts, so we can consider it as an L-type supercycle). The negative fractional period excess is poorly defined on the empirical diagram  "orbital period -- $\epsilon^{-}$" \citep{2001cvs..book.....H} for the shortest orbital periods. The extrapolation of the data suggests an expected $\epsilon^{-}$ in a region from -0.008 to -0.02. In the case of the orbital frequency $F_{orb} = 17.536$, a frequency of variations in quiescence $ F_{q} = 17.302$ could be a beat (side-band) frequency $F_{sb} = 2F_{orb} - F_{n}$ for $\epsilon^{-} = -0.013$ and the frequency of negative superhump $F_{n} = 17.767$.  

\subsection{Color \textbf{indices}}
\begin{figure}
   	\centerline{\includegraphics[width=0.95\textwidth,clip=]{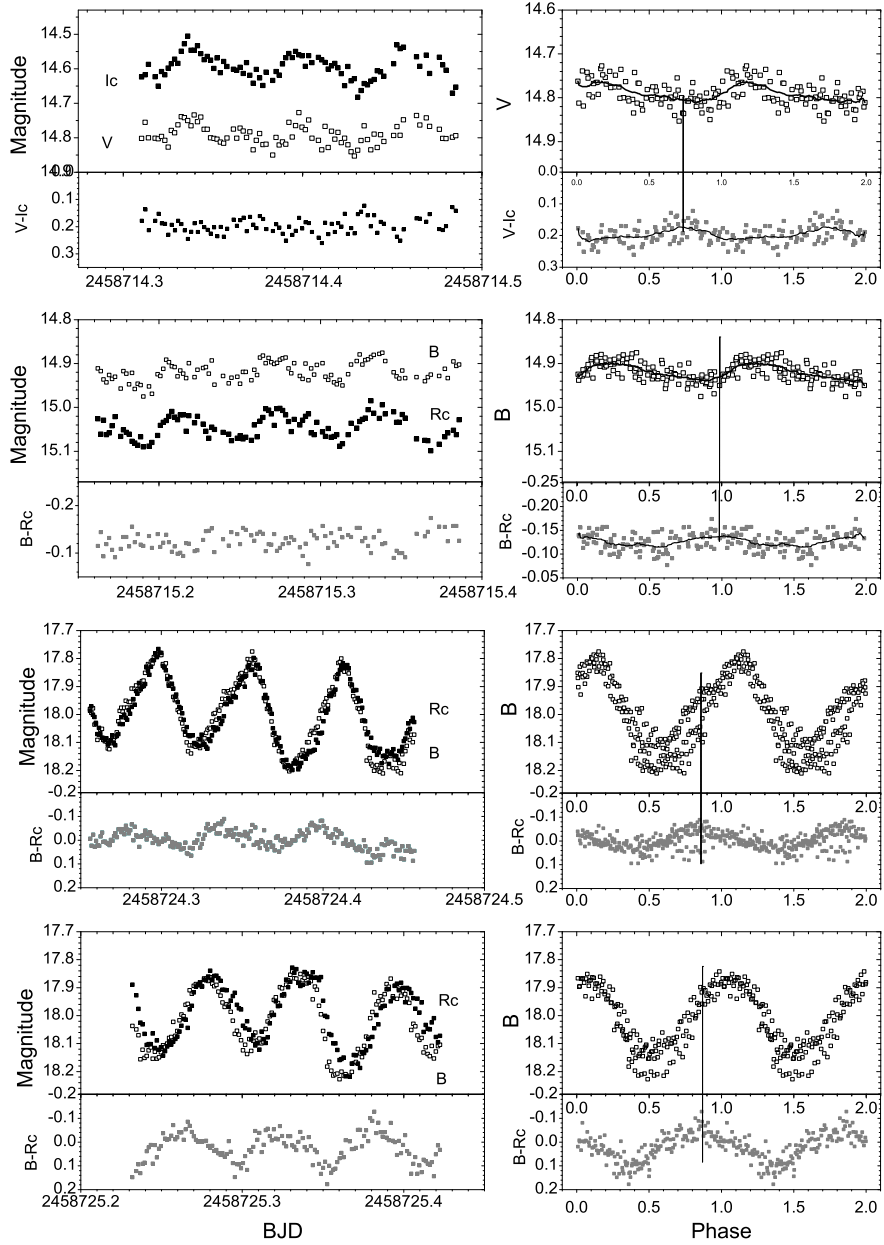}}
	\caption{Left from top to bottom: original light and colour indices curves for the data of the S2 superoutburst plateau (BJD 2458714 and 2458715 ) and two consecutive nights close to quiescence (BJD 2458724 and 2458725). Right from top to bottom: folded light (\textit{V} and \textit{Ic}, \textit{B} and \textit{Rc}) and instrumental colour indices (\textit{V-Ic} and \textit{B-Rc}) on the 0.058029 d period with the zero epoch BJD 2458724.0.  The dotted line is plotted through the \textit{B-Rc} maximum. The smoothed line is drawn through the \textit{V, V-Ic}, and \textit{B, B-Rc} light and colour indices curves, respectively.}
	\label{Fig7b}
\end{figure}

The multicolour \textit{V, Ic} and \textit{B, Rc} observations of superhumps were done for two nights of the S2 superoutburst plateau and those of \textit{B, Rc} were done during the slow  S2 superoutburst decline close to quiescence. As it follows from  Fig. \ref{Fig7b}, colour indices are the phase-dependent ones and their dependence is different for different stages of outburst activity. During the S2 superoutburst, the bluest peak of the \textit{V-Ic} and \textit{B-Rc} coincided with a maximum of the light curve, while during a slow approach to quiescence it was slightly shifted and coincided with a rising branch of the light curve. The amplitudes of colour curves were $\sim 0^{m}.08$ during the superoutburst and $\sim 0^{m}.1 - 0^{m}.2$ close to quiescence. Similar behaviour of the superhumps colour indices was reported for several SU UMa stars (see a brief review by \citet{2020Ap....tmp...67P}. 

\subsection{Mass ratio from photometry}

With known orbital and positive superhumps periods, one could calculate the fractional period excess $\epsilon^{+}$ at stage B of the superoutburst 
\begin{equation}
	\label{r1}
\epsilon^{+} = (Psh - Porb)/Porb,
\end{equation}  
where $Psh$ and $Porb$ are positive superhump and orbital periods,  $q=M_2/M_1 $, $M_2$ and $M_1$ are the masses of the donor and the primary components, respectively. We obtained $\epsilon^{+} = 0.0173$ for the stage B.

With a known orbital period and $\epsilon$ we can estimate the mass ratio  $q(\epsilon^{+})$. 
According to \citet{2013PASJ...65..115K}, the real $q$ one could obtain with the period of superhumps that is detected at stage A of the superoutburst, which is unknown in our case. As we pointed in Section 5, stage A was not recorded in any of the superoutbursts: it is seen from  Fig. \ref{fig3a} and Fig. \ref{fig4a} that the onset of the superoutburst is already accompanied with fully grown superhumps. The only thing that can be said is that a potential stage A lasted no more than a day. However, we could estimate expected $\epsilon^{*}$ at stage A using the relation of \citet{2013PASJ...65..115K},
\begin{equation}
\label{r2}
\epsilon^{*} = 0.012(2) + 1.04(8)\epsilon^{+},
\end{equation} 
where $\epsilon^{*}$ and $\epsilon^{+}$ are the fractional period excesses at stage A and B, respectively. We found $\epsilon^{*}$ = 0.030. Then, according to  Table 1  from \citet{2013PASJ...65..115K}, we obtain $q = 0.081$ and assuming $M_{1} = 0.75M_{\sun}$, we have $M_{2} = 0.061M_{\sun}$. 

\begin{figure}
	\centerline{\includegraphics[width=0.65\textwidth,clip=]{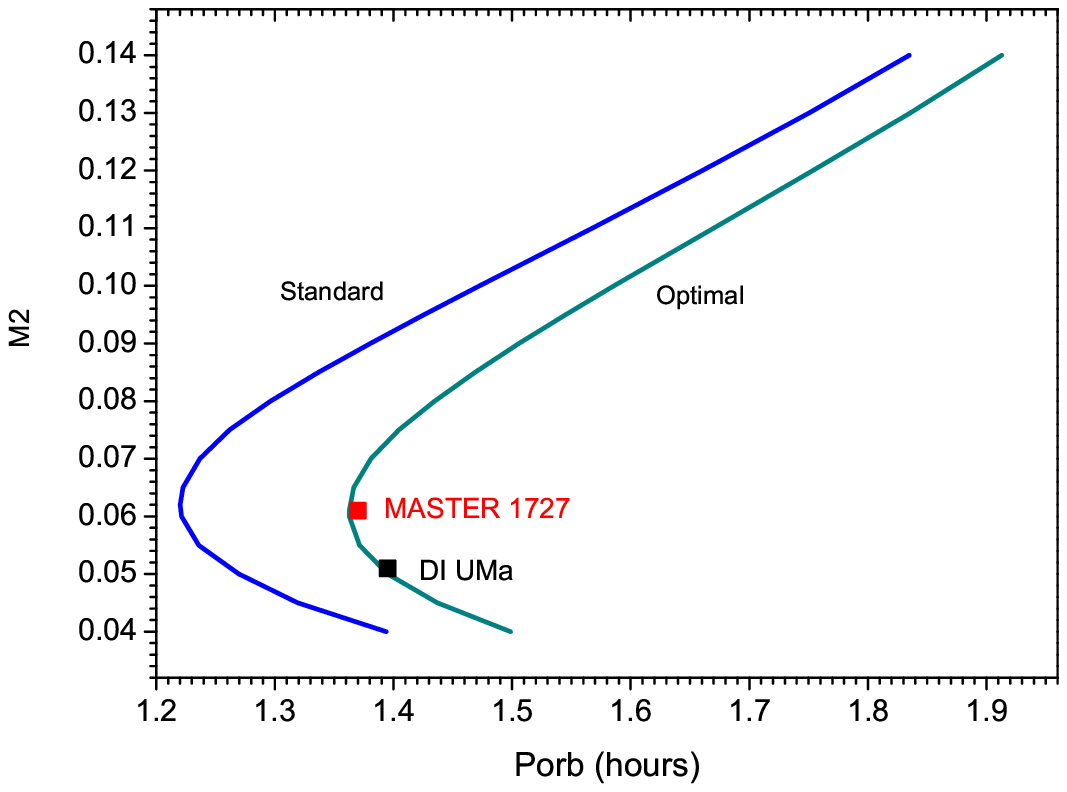}}
	\caption{Dependence of the mass of the secondary $M_{2}$, expressed in units of the solar masses, on the orbital period. For DI UMa we used $P_{orb}$ = 0.054564 d \citep{1999PASP..111.1275F} and re-estimated $q = 0.068$ and $M_2 = 0.051M_{\sun}$ obtained by the same method as for MASTER 1727. Tracks of standard and optional evolution according to \citet{2011ApJS..194...28K} are marked by blue and dark cyan lines, respectively. The position of DI UMa and MASTER 1727 is shown.}
	\label{knigge2a}
\end{figure}

This result is in a good agreement with the estimate for the mass described in Section 3 and places MASTER 1727 near a low-mass bound of the H-rich SU UMa stars (Fig. \ref{knigge2a}).

\section{Discussion: MASTER 1727 as a peculiar ER UMa-type star}

\begin{figure}
	\centerline{\includegraphics[width=0.85\textwidth,clip=]{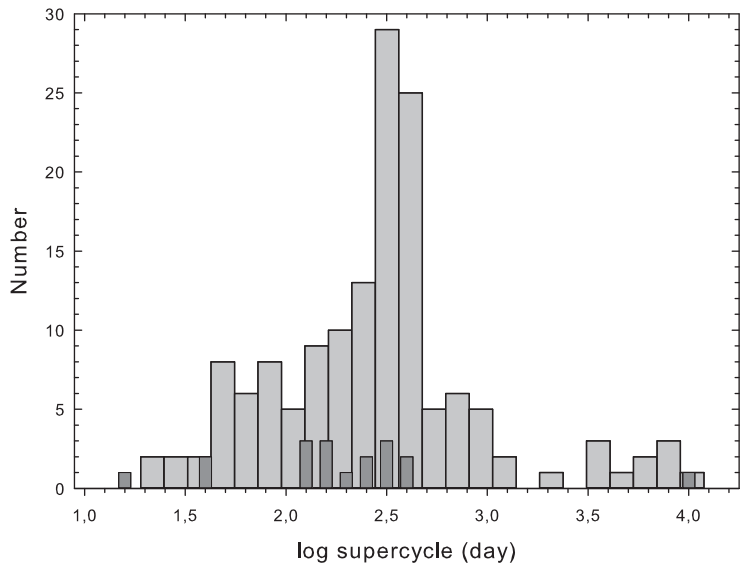}}
	\caption{A histogram of supercycles distribution for the SU UMa-type CVs based on the avalable  data taken from \citet{2003A&A...404..301R}, edition 7.24; \citet{2019cwdb.confE..39P}; \citet{2013Ap.....56..539A}.  Data from \citet{1999dicb.conf...45K} are marked by dark rectangles.}
	\label{fighist-2}
\end{figure}

In the 90s the known supercycles of SU UMa-type stars were mainly between $\sim$100 and $\sim$500 days \citep{1999dicb.conf...45K}. A special group of four objects with supercycles of 19 -- 44 d, called ER UMa-type stars \citep{2003A&A...404.1067N}, stood out among them. At that time, no new  SU UMa type-stars were known between these two groups. Over time, information began to appear  about both new stars of the ER UMa-type and stars with supercycles less than 100 d ("active novae"). Moreover, as it turned out, ER UMa itself displayed a change of its supercycle \citep{2013PASJ...65...54Z} in a range of 42-60 d. A modern histogram of the supercycles of SU UMa-stars distribution is presented in Fig. \ref{fighist-2}. This distribution is a bimodal one. The largest number of supercycles falls on 250-400 days (the primary maximum). A smaller, almost flat secondary maximum falls on the interval of 40-160 days and ER UMa-type stars no longer look like a separate group in this distribution. 

A short $\sim$50-d supercycle length could define MASTER 1727 as an ER UMa-type DN. Also, like other ER UMa-type stars, MASTER 1727 displays amplitudes of superhumps that are largest in the earliest stage of the superoutburst; similarly to RZ LMi \citep{1995PASJ...47L..25O}, it has an extremely short superoutburst duration. 

Besides of MASTER 1727, there are two other known ER UMa-type objects around the period minimum,  RZ LMi \citep{1995PASJ...47..897N} and DI UMa \citep{1996PASJ...48L..93K}. MASTER 1727 is the next after DI UMa object on the standard CV evolutionary path. The evolutionary state of RZ LMi may be different \citep{2016PASJ...68..107K}. 

According to light curve simulations based on the thermal-tidal instability model of SU UMa stars  \citep{1995PASJ...47L..11O}, the ER UMa-type stars have a high rate of mass transfer, about $\sim$ 10 times higher than that expected from the CV evolution. 
In our case, for the shortest supercycle of about 50 d, two possibilities  correspond to the mass-transfer rate of 4 or 7 {\textit{\.{M}}},  where {\textit{\.{M}}}  is the mass-transfer rate in units of $10^{16}$ $gs^{-1}$ for the disk radius 0.35a, where a is the binary separation.

The low mass ratio (and, hence, a small mass of the secondary) is a key parameter that may explain the peculiarities of MASTER 1727.
  A small mass of the secondary causes the weaker tidal torques compared to the usual SU UMa-type stars  \citep{1995PASJ...47L..25O}. The weak tidal torques lead to an increase in the radius of the accretion disc at the end of the superoutburst and a reduction in its duration since a smaller angular momentum will be removed during the superoutburst. This could explain a shortening of a supercycle because it takes a shorter time to replenish the angular momentum reservoir in the disc. \citet{1995PASJ...47L..25O} showed that dependence of a number of normal outbursts on supercycle for ordinary and ER UMa-type DNe (i.e. for DNe with low and high $\dot M$) is the opposite: while this number is inversely proportional to the supercycle for stars with low and moderate $\dot M$, it is proportional to the supercycle for stars with high $\dot M$.
 
   A big radius of the disk could provide 
 the possibility of accumulation of cold matter in the outer parts of the accretion disc beyond the 3:1 resonance that may be the cause of rebrightenings.  Note that the colour indices behaviour of MASTER 1727 during the superoutburst is also  similar to that in WZ Sge-type stars (see e.g. \citet{2018PASJ...70L...4I, 2017MNRAS.467..597N}). A strong variations in $\dot M$ probably may modify a morphology of the superoutburst from a "clear" superoutburst to a superoutburst with rebrightenings.
   
 The supercycle variation of MASTER 1727 deserves special attention. According to \citet{1995PASJ...47L..25O}, a two-fold increase in the supercycle corresponds to a $\sim$ two-fold decrease of $\dot M$. There are several known examples with variable supercycles in active novae where the supercycle changes in a form of secular increase and/or quasi-periodic oscillations as was shown by  \citet{2001PASJ...53L..17K}, \citet{2014ASPC..490..385O}, \citet{2013PASJ...65...54Z}. Generally, the nature of these variations is still an open question. However, the tendency of the secular increase in the supercycle, which means a gradual increase in $\dot M$, suggests that these objects (at least some of them) could experience a transition between DNe and nova-likes. Thus \citet{2016PASJ...68..107K} applied this idea for a systematic increase of the supercycle length in RZ LMi. 
 
 There are several known postnovae stars of different evolution status (i.e., orbital periods)  showing some activity that is probably caused by a high mass transfer rate after the nova explosion. Thus in the renown classical  Nova Cygni 1975 (V1500 Cyg), which  is an asynchronous polar with the orbital period of 0.14d, a strong eruption heated the WD up to the $\sim$100 000 K \citep{1995ApJ...441..414S}. The hot WD irradiated the secondary and formed a radiation-driven wind \citep{1991ApJ...374L..59K}, which during the first  three years after the eruption could have an even stronger decelerating effect on the orbital motion of the WD than the magnetic field. The observational evidence of this was shown by \citet{1995ASSL..205..172P}. At present (45 years after the WD eruption) the white dwarf is still hot and produces $\sim$1.5-mag reflection effect of the secondary \citep{2018MNRAS.479..341P}.
 
 It was shown by \citet{2020arXiv201007812P}  that the mass-transfer rate in recurrent novae T Pyx and IM Nor is unnaturally high for their state of evolution (orbital periods are 0.076 d and 0.103 d, respectively) and supposed that it is caused by a strong and long-lasting irradiation effect of the secondary component  powered by a WD eruption.

 MASTER 1727, with an orbital period of 0.057 d  and mass ratio 0.08  is a low-q extension of  ER UMa stars. The mass-transfer rate should be accordingly higher than what is expected from gravitational radiation only.  The only currently known mechanism could be a post-nova state. They suggest that the object underwent a nova eruption relatively recently  -- hundreds of years ago. (Note that checking the DASCH archive\footnote{http://dasch.rc.fas.harvard.edu/} did not reveal a nova outburst over the past $\sim$100 years).
 This object would provide the evidence that a nova eruption can occur even if CVs are near the period minimum (probably the first observational evidence for that).
 
\section{Summary}

\begin{itemize}
\setlength\itemsep{-0.99em}
\item	The spectrum of MASTER J1727 that was obtained in 2020 quiescence contained H lines and no He lines;\\
\item	The continuous photometry in 2019 over $\sim $ 160 d  revealed a supercycle length to be 52 d, but the supercycle variability was found from 50 to 100 d on the $\sim$5.5-year scale. This implies as high as two times variations in the mass ratio; \\ 
\item	We found a mean period of positive superhumps to be 0.058029 d during a rather short 7-d superoutburst plateau;\\
\item	Late superhumps with a slightly shorter period of 0.057915 d were detected after the superoutburst. They lasted about 20 d up to the start of a normal outburst;\\
\item	In 2019 we found the orbital period of 0.057026 d during quiescence after the normal outburst and established its identity with those found earlier by spectroscopy \citep{2016AJ....152..226T};  the absence of an eclipse in the orbital light curve and its moderate amplitude is consistent with the orbital inclination of about 40$\degr$ that we found from spectral observations;\\
\item	A blue peak of \textit{V-Ic} and \textit{B-Rc} colour indices coincides with a minimum of the superhump light curve during the superoutburst and those of \textit{B-Rc} colour-index coincides with a rising branch of the late superhumps in quiescence;\\
\item  We estimated a low mass ratio of 0.08 and a small mass of the secondary $\sim$0.06 M$\sun$ that could explain most of the observed peculiarities of MASTER J1727 in the frame of thermal-tidal instability of accretion discs;\\ 
\item The high mass-transfer rate indicates that in the relatively recent history, the MASTER J1727  could undergo a nova eruption. If so, it means that an eruption can occur even if the star is near the period minimum. 
\end{itemize}

\acknowledgements
We are grateful to the anonymous referee for the valuable comments
which helped to improve this paper.
Elena Pavlenko is grateful to Nikolaus Vogt for the valuable discussion about
the MASTER 1727 at the "Compact White Dwarf Binaries" Conference held in Erevan (2019) and Natali Katysheva for careful reading of the manuscript.
Observations with the SAO RAS telescopes are supported by the Ministry of Science and Higher Education of the Russian Federation (including agreement No 05.619.21.0016, project ID RFMEFI61919X0016); Observations with CrAO RAS telescopes are supported by the grant of RSF (project number 19-72-10063).
The work of Shimansky was funded by RFBR, project number 18-42-160003.
The work of Maksim Gabdeev was funded by the subsidy 671-2020-0052 allocated to the Kazan Federal University for the state assignment in the sphere of scientific activities and was funded by RFBR, project number 19-32-60021.
The work of Dubovsk\'{y} was supported
by the Slovak Research and Development Agency under contract No.
APVV-15-0458.

\bibliography{pavlenko_caosp307}

\end{document}